\documentclass[twocolumn]{aastex63}

\usepackage[T1]{fontenc}
\usepackage[utf8]{inputenc}
\usepackage[letterspace=-15]{letterspace}

\received{October 22, 2020}
\revised{November 6, 2020}
\accepted{November 17, 2020}
\submitjournal{ApJL}

\shorttitle{A wide planetary-mass companion in Ophiuchus}
\shortauthors{Fontanive et al.}
\graphicspath{{./}{figures/}}

\begin{document}

\title{A wide planetary-mass companion to a young low-mass brown dwarf in Ophiuchus}

\correspondingauthor{Clémence Fontanive}
\email{clemence.fontanive@csh.unibe.ch}

\author[0000-0002-2428-9932]{Clémence Fontanive}
\affiliation{Center for Space and Habitability, University of Bern, Gesellschaftsstrasse 6, 3012 Bern, Switzerland}

\author[0000-0003-0580-7244]{Katelyn N. Allers}
\affiliation{Department of Physics and Astronomy, Bucknell University, Lewisburg, PA 17837, USA}

\author{Blake Pantoja}
\affiliation{Department of Physics and Astronomy, Bucknell University, Lewisburg, PA 17837, USA}

\author[0000-0003-4614-7035]{Beth Biller}
\affiliation{SUPA, Institute for Astronomy, University of Edinburgh, Blackford Hill, Edinburgh EH9 3HJ, UK}

\author{Sophie Dubber}
\affiliation{SUPA, Institute for Astronomy, University of Edinburgh, Blackford Hill, Edinburgh EH9 3HJ, UK}

\author[0000-0002-3726-4881]{Zhoujian Zhang}
\affiliation{Institute for Astronomy, University of Hawai’i, 2680 Woodlawn Drive, Honolulu, HI 96822, USA}

\author[0000-0001-9823-1445]{Trent Dupuy}
\affiliation{SUPA, Institute for Astronomy, University of Edinburgh, Blackford Hill, Edinburgh EH9 3HJ, UK}

\author[0000-0003-2232-7664]{Michael C. Liu}
\affiliation{Institute for Astronomy, University of Hawai’i, 2680 Woodlawn Drive, Honolulu, HI 96822, USA}

\author[0000-0003-0475-9375]{Loïc Albert}
\affiliation{Institut de recherche sur les exoplanètes, Université de Montréal, Montréal, H3C 3J7, Canada}



\begin{abstract}

We present the discovery of a planetary-mass companion to CFHTWIR-Oph~98, a low-mass brown dwarf member of the young Ophiuchus star-forming region, with a wide 200-au separation (1\farcs46). The companion was identified using \textit{Hubble Space Telescope} images, and confirmed to share common proper motion with the primary using archival and new ground-based observations. Based on the very low probability of the components being unrelated Ophiuchus members, we conclude that Oph~98~AB forms a binary system.
From our multi-band photometry, we constrain the primary to be an M9--L1 dwarf, and the faint companion to have an L2--L6 spectral type. For a median age of 3~Myr for Ophiuchus, fits of evolutionary models to measured luminosities yield masses of $15.4\pm0.8$~M$_\mathrm{Jup}$ for Oph~98~A and $7.8\pm0.8$~M$_\mathrm{Jup}$ for Oph~98~B, with respective effective temperatures of $2320\pm40$~K and $1800\pm40$~K. For possible system ages of 1--7~Myr, masses could range from 9.6--18.4~M$_\mathrm{Jup}$ for the primary, and from 4.1--11.6~M$_\mathrm{Jup}$ for the secondary.
The low component masses and very large separation make this binary the lowest binding energy system imaged to date, indicating that the outcome of low-mass star formation can result in such extreme, weakly-bound systems. With such a young age, Oph~98~AB extends the growing population of young free-floating planetary-mass objects, offering a new benchmark to refine formation theories at the lowest masses.

\end{abstract}



\section{Introduction} \label{sec:intro}

\begin{figure*}
    \centering
    \includegraphics[width=\textwidth]{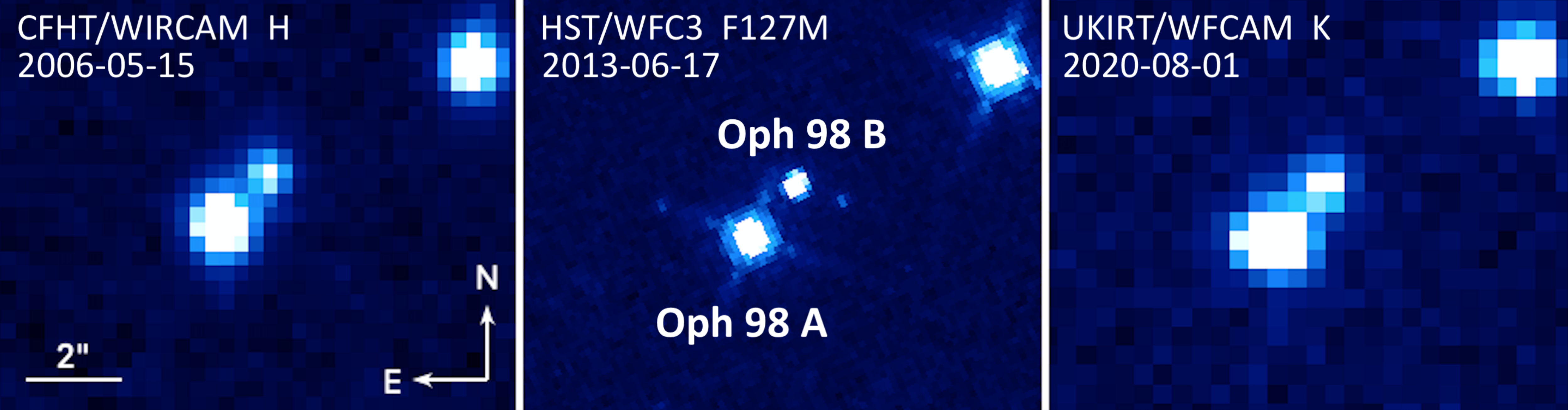}
    \caption{Images of the Oph~98 binary system from the first epoch of CFHT data (left), \textit{HST} observations (middle) and UKIRT data (right). North is up and East is left. The angular scale is indicated in the left panel and is the same for all images. The source North-West of the binary (upper right corner) is the reference background star \textit{Gaia}~DR2~6050679111185185664.}
    \label{fig:Oph98-image}
\end{figure*}

Currently only a handful of planetary-mass companions ($<13$~M$_\mathrm{Jup}$) are known around young ($<20$~Myr) brown dwarfs \citep{Chauvin2005,Todorov2010,Bejar2008,Best2017,Dupuy2018}. These objects are unlikely to have formed in the disk of their primary, however, they possess similar masses and temperatures as young self-luminous giant exoplanets.
This small but growing population of binaries provides critical tests for theoretical models. While the method of their formation differs from that of planets orbiting stars, the frequency and properties of such systems constrain formation theories of the lowest-mass objects. With well-determined ages compared to field brown dwarfs, each of these binaries also provides an important archetypal system for validating atmosphere and evolutionary models.  

The best studied of these companions, 2M1207b \citep{Chauvin2005}, was quite puzzling at the time of its discovery, as it appears to have very red near-infrared colors compared to model predictions given its luminosity \citep{Mohanty2007}. Interestingly, the first directly-imaged exoplanets are similarly red \citep{Marois2008,Barman2011}. This is attributed to the low surface gravity of young objects and is found as well for free-floating brown dwarfs (e.g. \citealp{Cruz2009,Faherty2009,Liu2016}). Low surface gravity has implications for the cloud structure of these benchmark objects -- their red colors suggest that they retain dusty silicate clouds down to much lower effective temperatures compared to high surface gravity field brown dwarfs at similar temperatures. Discovering and characterizing even younger planetary-mass objects allows us to study these atmospheres at the very youngest ages and lowest surface gravities. 

2M1207b is a member of the $\sim$11-Myr TW Hya association; identifying similar systems at even younger ages enables strong constraints on the formation of such objects (cluster environment, accretion processes, presence of disks), as well as to trace the early evolution of their physical properties.
In this Letter, we report the discovery of such a young low-mass binary, CFHTWIR-Oph~98 (hereafter Oph~98; 2MASS J16274422$-$2358521). The brown dwarf Oph~98~A is a $\sim$15-M$_\mathrm{Jup}$ member of the $\sim$3-Myr Ophiuchus star-forming region \citep{AlvesdeOliveira2012}. Using ground- and space-based observations, we confirm and characterize Oph~98~B as a faint and red co-moving companion with a mass of $\sim$8~M$_\mathrm{Jup}$, and a large separation of about 200~au.

\section{Observations and Data Analysis} \label{sec:observations}

\subsection{Hubble Space Telescope data} \label{sec:HST}

We observed Oph~98 as part of a \textit{Hubble Space Telescope} (\textit{HST}) multiplicity survey (GO 12944, PI Allers) targeting brown dwarfs in the Ophiuchus star-forming region. Data for this target were acquired on UT 2013 June 17. Two sets of deep dithered images were obtained in the F127M and F139M filters on the IR channel of the WFC3 instrument, in full frame MULTIACCUM mode, with total exposure times of 698~s in each band. Two images of 473~s each were then acquired in the F850LP bandpass on the UVIS channel, using the full UVIS aperture. 
The combination of these three filters allows for clear distinctions of sub-stellar objects from background interlopers, by exploiting the inherent red colors of brown dwarfs and a characteristic water absorption band observed in their spectra at 1.4~$\mu$m \citep{Fontanive2018,Allers2020}. Oph~98~A was indeed found to show distinctive photometric colors between these bands compared to other stars in the \textit{HST} field of view. A faint companion, well resolved in all images and shown in the F127M band in Figure~\ref{fig:Oph98-image} (middle panel), was detected at $\sim$1\farcs46 from the known brown dwarf based on its multi-band photometry.

The pipeline processed flat-field images were used as input in the {\tt MultiDrizzle} software \citep{FruchterHook2002} to correct for geometric distortion, perform cosmic ray rejection and combine all dithered frames into a final image in each filter. Source positions were extracted using the {\tt DAOStarFinder} algorithm from the {\tt Photutils} python package \citep{Bradley2019}. Aperture photometry was performed adopting 0\farcs4 aperture radii. The background level and its uncertainty in each final data frame was estimated by applying the same 0\farcs4 aperture to 2000 random star-free positions and computing the mean and standard deviation of these measurements. Measured fluxes were finally converted into Vega magnitudes using the \textit{HST} photometric zero points for 0\farcs4 apertures in the considered filters.
Measured magnitude differences ($\Delta$mag) and relative positions between Oph~98~A and B are reported in Table~\ref{t:measurements}. Apparent \textit{HST} magnitudes are listed in Table~\ref{t:properties}.

\begin{table*}
    \centering
    \caption{Photometric and astrometric measurements of Oph 98 AB}
    \footnotesize
    \begin{tabular}{l c c c c c}
    \hline \hline
        UT Date & Telescope/Instrument & Filter & $\Delta$mag & Separation & Position Angle  \\
        &  &  &  & [mas] & [deg] \\
        \hline
        2006 May 16 & CFHT/WIRCAM & \textit{J} & $2.089\pm0.032$ & $1503.67\pm19.94$ & $318.0\pm0.8$ \\
        2006 May 15 & CFHT/WIRCAM & \textit{H}\textit{} & $1.760\pm0.055$ & $1451.31\pm26.4$ & $318.7\pm1.1$ \\
        2006 May 16 & CFHT/WIRCAM & \textit{Ks} & $1.536\pm0.020$ & $1478.6\pm15.11$ & $319.7\pm0.6$ \\
        2012 Aug 10 & CFHT/WIRCAM & \textit{J} & $2.169\pm0.077$ & $1478.76\pm33.47$ & $319.0\pm1.3$ \\
        2012 Aug 10 & CFHT/WIRCAM & \textit{H} & $1.755\pm0.033$ & $1485.08\pm17.78$ & $318.8\pm0.7$ \\
        2012 Aug 10 & CFHT/WIRCAM & \textit{Ks}\textit{} & $1.630\pm0.021$ & $1443.32\pm14.99$ & $319.6\pm0.6$\\
        2013 Jun 17 & \textit{HST}/WFC3 & F850LP & $2.783\pm0.171$ & $1458.04\pm1.54$ & $318.8\pm0.1$ \\
        2013 Jun 17 & \textit{HST}/WFC3 & F127M & $2.192\pm0.014$ & $1458.19\pm6.94$ & $319.1\pm0.3$ \\
        2013 Jun 17 & \textit{HST}/WFC3 & F139M & $2.032\pm0.017$ & $1460.55\pm6.46$ & $319.2\pm0.3$ \\
        2020 Aug 01 & UKIRT/WFCAM & \textit{J} & $2.073\pm0.028$ & $1462.58\pm20.21$ & $316.7\pm0.8$ \\
        2020 Aug 01 & UKIRT/WFCAM & \textit{K} & $1.642\pm0.039$ & $1435.07\pm8.49$ & $322.0\pm0.4$ \\
        2020 Sep 17 & UKIRT/WFCAM & \textit{H} & $1.793\pm0.025$ & $1519.51\pm21.57$ & $319.0\pm0.8$ \\
         \hline
    \end{tabular}
    \label{t:measurements}
\end{table*}

\subsection{Canada-France-Hawaii Telescope data} \label{sec:CFHT}

Seeing-limited, broad-band \textit{J}, \textit{H}, and \textit{Ks} (MKO filter system) images of Oph~98 are available from the Canada-France-Hawaii Telescope (CFHT) WIRCAM archive (Programs 06AF01, 06AT08, and 12AT09), with data acquired in May 2006 and August 2012. A CFHT \textit{H}-band image of Oph~98 is shown in left panel of Figure~\ref{fig:Oph98-image}.

We used pre-processed images from the CFHT facility pipeline, 'I'iwi, which includes detrending and sky subtraction. 
We determined the relative astrometry and photometry of Oph~98~A and B from individual pre-processed frames.  We used the IDL Astronomy Library's {\tt FIND}, {\tt APER}, and {\tt GETPSF} routines \citep{IDLAstro} to determine the point spread function (PSF) and residuals for non-saturated stars within 1~arcmin of Oph~98~A.  Using the {\tt NSTAR} program, we fit the PSF to Oph~98~A, and subtracted it from the image using {\tt SUBSTAR}.  We then fit a model PSF to Oph~98~B.  For each epoch and each filter, we determined the separation, position angle, and $\Delta$mag of the binary using the mean and standard deviation of the mean of measurements from individual (non-stacked) frames.  Table~\ref{t:measurements} reports our measurements.

We also determined MKO photometry of Oph~98~A and B from the individual, pre-processed frames.  We used the magnitudes calculated by {\tt NSTAR} during the PSF fitting process, and determined the photometric calibration offset for each image by comparing the PSF-fit magnitudes of non-saturated stars within 1~arcmin of Oph~98~A to their 2MASS photometry. We first converted their 2MASS magnitudes to the MKO system using custom color corrections derived from synthetic photometry of SpeX Spectral Library spectra reddened by $A_V$ of 1--30~mag.  We calculated the photometry for Oph~98~A and B using a weighted mean and weighted standard deviation of the mean of the magnitudes calculated from each frame.  Our photometry (Table~\ref{t:properties}) is in good agreement with published WIRCAM photometry of Oph~98~A from \citet{AlvesdeOliveira2012}.

\subsection{United Kingdom Infra-Red Telescope data} \label{sec:UKIRT}

We obtained seeing-limited \textit{J} and \textit{K}-band images of Oph~98 using the United Kingdom Infra-Red Telescope (UKIRT) WFCAM instrument on UT 2020 August 01, followed by an \textit{H}-band dataset on UT 2020 September 17 (Project ID U/20A/H02). The \textit{K}-band image is shown in Figure~\ref{fig:Oph98-image} (right panel).  Using the same procedure described in \S~\ref{sec:CFHT}, we determined the photometric and astrometric measurements of Oph~98~AB from individual, pre-processed images, reported in Tables~\ref{t:measurements} and~\ref{t:properties}.

\section{Characterization of the Binary} \label{sec:charac}

\subsection{Astrometric Analysis} \label{sec:astrometry}

\begin{figure*}
    \centering
    \includegraphics[width=0.4\textwidth]{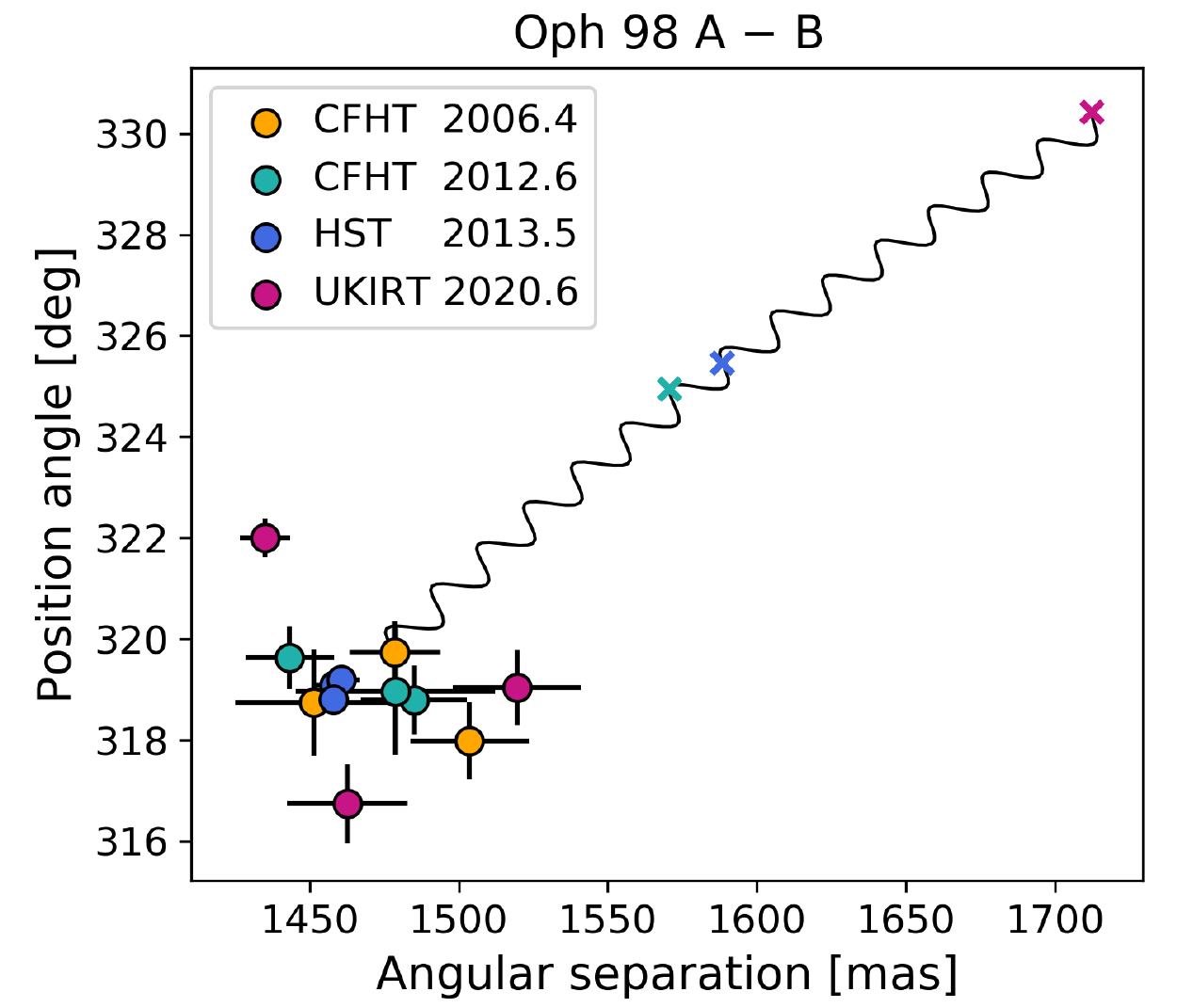}
    \hspace{0.6cm}
    \includegraphics[width=0.4\textwidth]{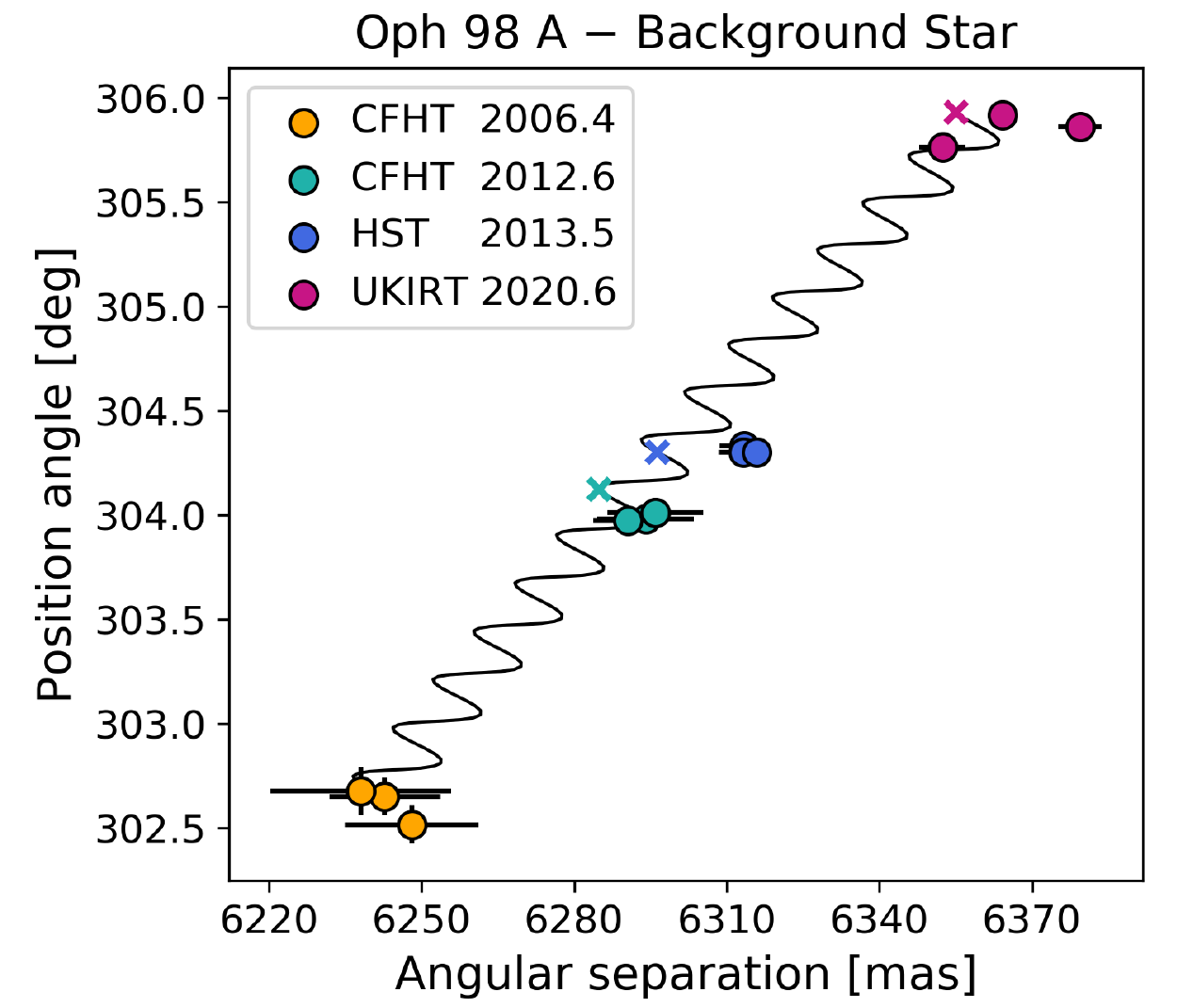}
    \caption{Positions of Oph~98~B (left) and the nearby star \textit{Gaia}~DR2 6050679111185185664 (right) relative to Oph~98~A. Measurements are indicated by filled circles, color-coded by observational epoch. The solid black lines show the expected motion of a stationary background star relative to the primary past the first observational epoch, given the parallax and proper motion of Ophiuchus (Table~\ref{t:properties}). Crosses mark the expected positions of a background source at the given color-coded dates. Oph~98~B is clearly co-moving with Oph~98~A, while the relative motion of the \textit{Gaia} star is fully consistent with a background source.}
    \label{fig:CPM}
\end{figure*}

The relative astrometry of the binary was measured in the various imaging data sets available as detailed in \S~\ref{sec:observations} (Table~\ref{t:measurements}). The left panel of Figure~\ref{fig:CPM} clearly demonstrates that Oph~98~A and B are co-moving over the 14-year time baseline of ground- and space-based data, indicating that the two components share a common proper motion. Based on their coordinates, Oph~98~A and B are most likely part of the young L1688 cloud in the Ophiuchus complex \citep{EsplinLuhman2020}. Given the stellar density of L1688 \citep{King2012}, the chance of alignment for two unrelated Ophiuchus members within 2\arcsec~is $<10^{-4}$. This number is an overestimate in the case of such rare low-mass brown dwarfs at the bottom of the initial mass function \citep{Kroupa2013}, and we conclude that Oph~98~A and B form a physically-associated binary pair.

A nearby star (\textit{Gaia}~DR2 6050679111185185664), $\sim$6\arcsec~North-West from Oph~98, was found to be in the \textit{Gaia} Data Release~2 (DR2) catalog \citep{Gaia2016,Gaia2018} with a full astrometric solution. With \textit{Gaia} parallax and proper motion measurements of $\varpi=1.453\pm0.641$~mas, $\mu_{\alpha*}=-0.795\pm1.396$~mas~yr$^{-1}$ and $\mu_{\delta}=-1.192\pm0.824$~mas~yr$^{-1}$, this source is essentially consistent with a stationary background star. As shown in the right panel of Figure~\ref{fig:CPM}, the positions of the star relative to Oph~98~A at each observational epoch are in excellent agreement with the expected relative displacement over time for a background star (black line). The small positional disparities are consistent with the almost negligible motion of the \textit{Gaia} source and with the measurement uncertainties in the original epoch from which the background track is calculated. These results confirm that the primary has a proper motion expected for Ophiuchus, hence validating the Ophiuchus membership of the co-moving Oph~98~AB system.

We adopt the parallax of $\varpi=7.29\pm0.22$~mas derived by \citet{Ortiz-Leon2018} for the embedded L1688 population based on \textit{Gaia}~DR2, in good agreement with values from \citet{Canovas2019} and \citet{EsplinLuhman2020}. At a corresponding distance of $137\pm4$~pc, the observed angular separation of the Oph~98 binary (1\farcs46$\,\pm\,$0\farcs01) implies a wide projected separation of $200\pm6$~au.

\subsection{Photometric Estimates of Spectral Types} \label{sec:SpT}

In order to estimate the spectral types of Oph~98~A and B, we used the SpeX Prism Library Analysis Toolkit (SPLAT; \citealp{SPLAT}) to fit the spectral energy distribution (SED) of the binary components. We gathered a library of M and L spectra from the SPLAT database, for which we obtained homogeneous near-infrared spectral types and gravity scores using the \citet{AllersLiu2013} classification, and retained only young sources with very-low (\textsc{vl-g}) gravity scores.

SPLAT allows for the determination of photometric magnitudes on specific filters based on a source's spectrum. The module can also redden a spectrum following the \citet{Cardelli1989} reddening law.
We used these capabilities to determine the scaling factor and visual extinction $A_V$ minimizing the $\chi^2$ between the synthetic photometry of the templates and our measured magnitudes for Oph~98~A and B. 
We added uncertainties of 0.011, 0.007 and 0.007~mag (2MASS calibration uncertainties; \citealp{2MASS}) to the \textit{JHK} MKO measurements tied to 2MASS, and 2\% and 5\% uncertainties in the \textit{HST} IR and UVIS photometry\footnote{\url{https://www.stsci.edu/hst/instrumentation/wfc3/data-analysis/photometric-calibration}}, respectively. 

We observed a strong degeneracy between spectral type and extinction in our results, consistent with findings by \citet{Luhman2017} for young reddened L dwarfs. Results for Oph~98~A showed a handful of fits with similar $\chi^2$ values for objects with spectral types of M9 to L1, and decreasing reddening values with later type over the range $A_V\!\sim\!4$--6~mag. A similar effect was seen in the results of Oph~98~B, but spanning wider ranges of spectral types (late-M to mid-L) and extinctions ($A_V\!\sim\!4$--10~mag), likely due to the atypical colors and larger uncertainties on the photometry of the secondary. Assuming the same local cloud extinction for the two components, we can reduce this range to best-fit results yielding $A_V\!<\!6$~mag for the companion. This provides spectral type estimates of M9--L1 for Oph~98~A, and L2--L6 for Oph~98~B, with a visual extinction for the system of $A_V=5\pm1$~mag, placing the primary at the M/L transition and loosely constraining the secondary to be along the L spectral sequence.
Our derived quantities for the primary are in reasonable agreement with the values of M9.75 and $A_V\!=\!3$~mag estimated by \citet{AlvesdeOliveira2012} from low-resolution $H$- and $K$-band spectroscopy. Additional spectroscopic observations will be required to further characterize the system and break down the observed degeneracies.

\begin{figure*}
    \begin{minipage}{0.52\textwidth}
    \centering
    \includegraphics[width=0.98\textwidth]{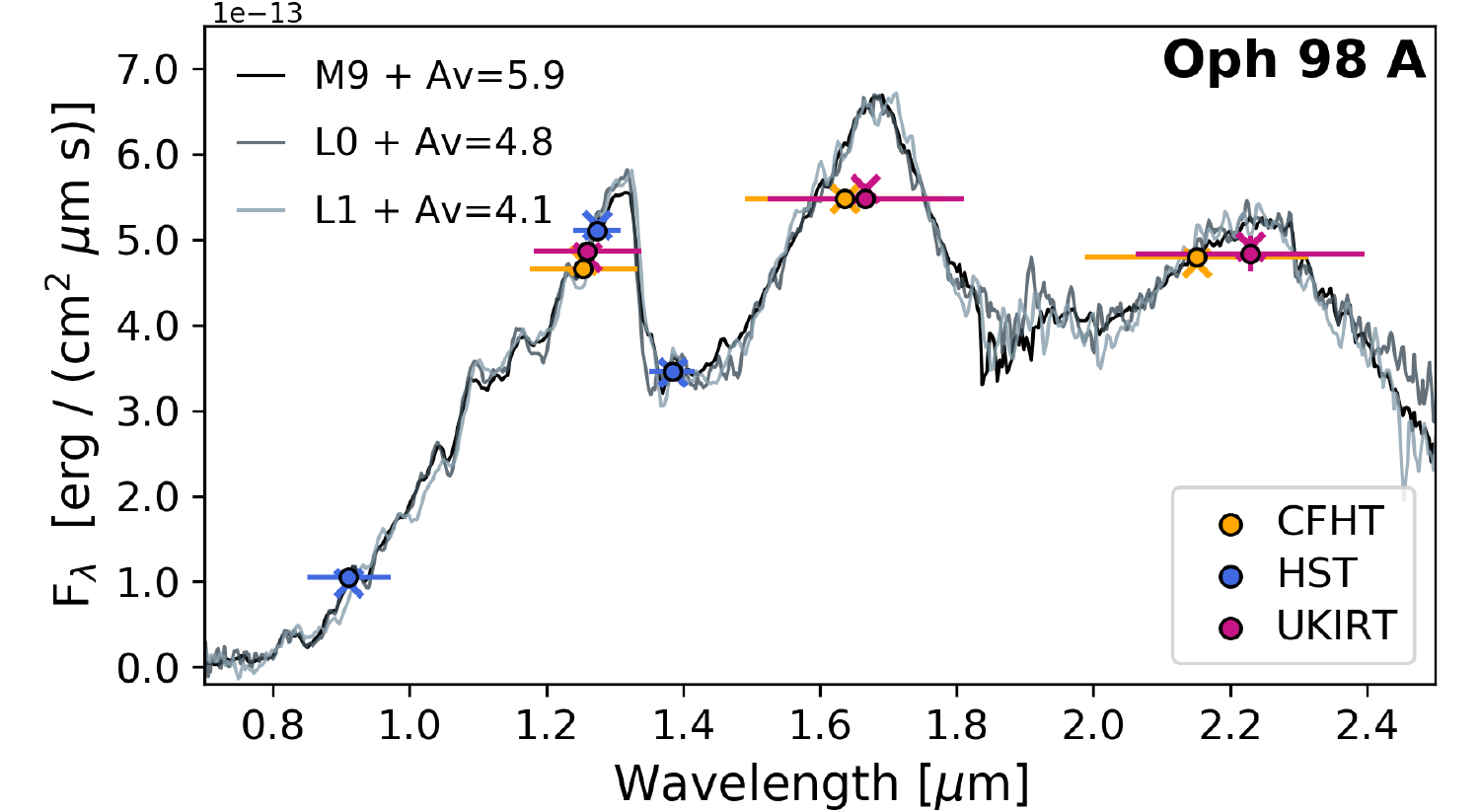}
    \includegraphics[width=0.98\textwidth]{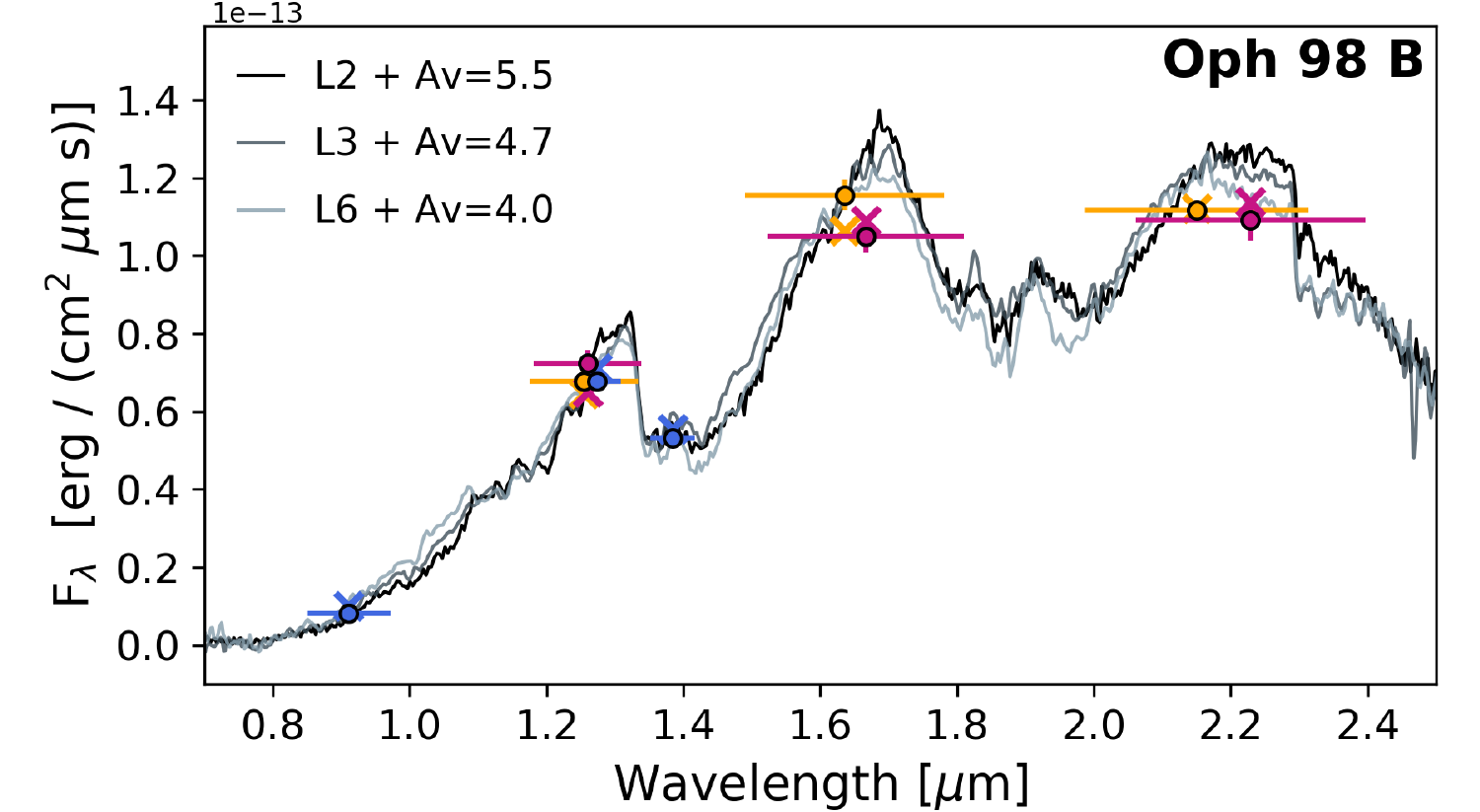}
    \end{minipage}
    \begin{minipage}{0.48\textwidth}
    \includegraphics[width=0.98\textwidth]{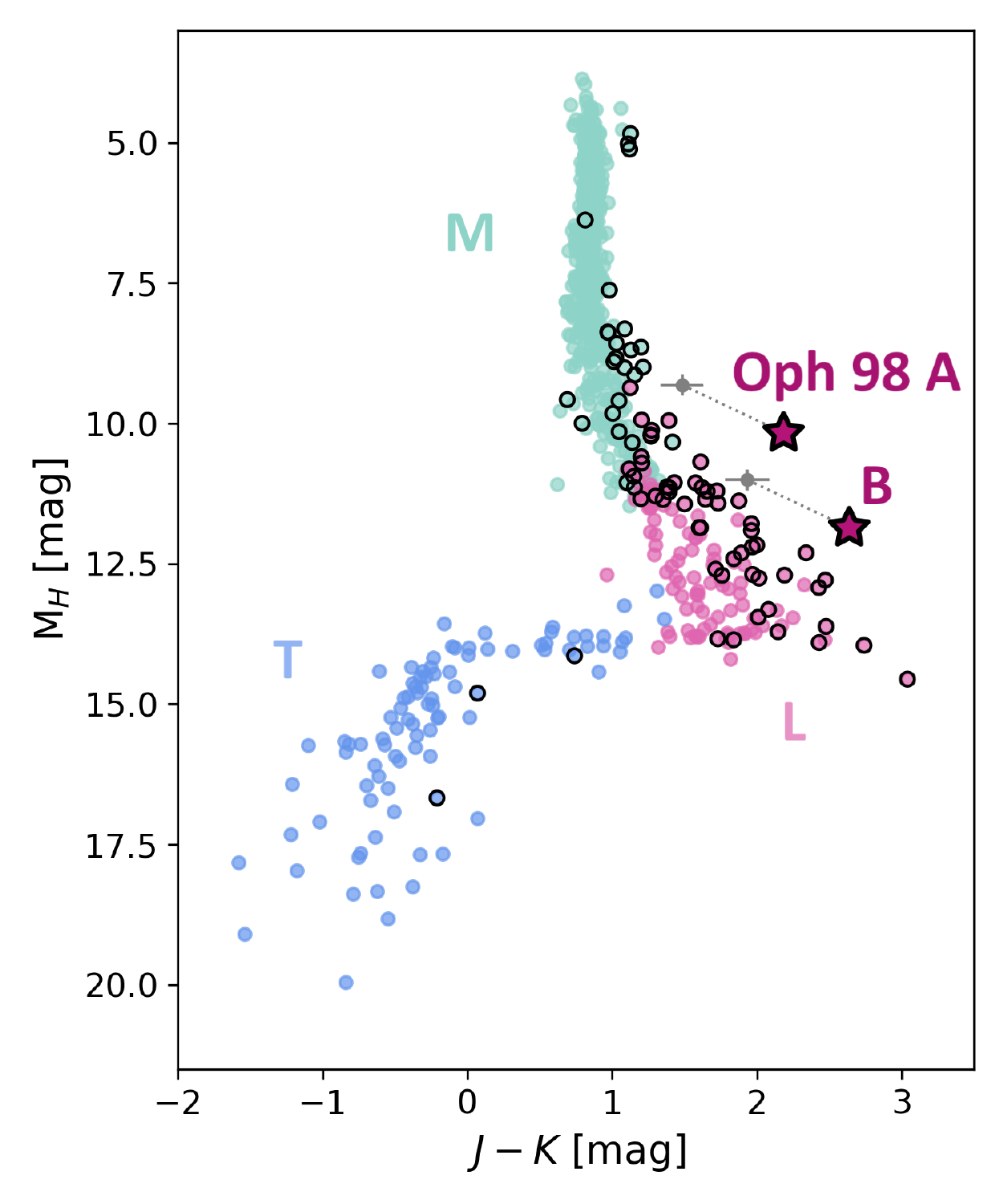}
    \end{minipage}
        \caption{\textbf{Left:} SED fits of Oph 98 A (top) and B (bottom). Photometric measurements (filled circles) are compared to a selection of best-fit templates, reddened by their fitted extinctions: TWA~26 (M9~\textsc{vl-g}), PSO~J078.9904$+$31.0171 (L0~\textsc{vl-g}), 2MASS~J16410015$+$1335591 (L1~\textsc{vl-g}), 2MASSW~J2206450$-$421721 (L2~\textsc{vl-g}), 2MASSW~J0030300$-$145033 (L3~\textsc{vl-g}), SDSSp~J010752.33$+$004156.1 (L6~\textsc{vl-g}). Crosses show the synthetic photometry in each band for the L0 and L3 spectra. \textbf{Right:} Near-infrared color-magnitude diagram of MLT objects showing \textit{H}-band absolute magnitudes against \textit{J}--\textit{K} colors. The observed WFCAM photometry of Oph~98~AB is plotted in the magenta stars, connected by the dotted lines to extinction-corrected values for $A_V=5\pm1$~mag (grey dots). Objects with black circles are young, low-gravity objects and directly-imaged young companions.}
    \label{fig:SED_CMD}
\end{figure*}

The left panel of Figure~\ref{fig:SED_CMD} shows best-fit template spectra for both components compared to our photometric measurements.
The right panel shows the observed positions of Oph~98~A and B (magenta stars) in the near-infrared color-magnitude diagram compared to the population of low-mass stars and brown dwarfs. M dwarfs come from \citet{Winters2015}. Late-M, L and T dwarfs are compiled from \citet{DupuyLiu2012}, \citet{DupuyKraus2013} and \citet{Liu2016}. Bold circles represent young objects and companions with low surface gravities, which extend the standard M and L sequences to redder \textit{J}--\textit{K} colors. The grey dots connected to Oph~98~A and B indicate the extinction-corrected locations of the binary for our derived extinction value. However, since the reddening is determined by best fits to templates, the dereddened colors may not be representative of the intrinsic colors of the components.

\subsection{Physical Properties} \label{sec:properties}

Since the age of Ophiuchus can be inferred from stellar members of the region, the luminosities of Oph~98~A and B can provide estimates of the physical properties of the binary using evolutionary models.  Recently, \citet{EsplinLuhman2020} analyzed the ages of various populations within the Ophiuchus star-forming complex. They estimated ages of $\sim$2 and $\sim$3--4~Myr for embedded and low-extinction members of L1688, respectively. Our estimated extinction for Oph~98 ($A_V=5\pm1$ mag) falls on the boundary of $A_V$ between the two L1688 populations defined by \citet{EsplinLuhman2020}. Thus, we adopt an age of 3~Myr for Oph~98, but ages of 1--7~Myr fall within the interquartile age ranges of L1688.

\begin{table*}
    \setlength\tabcolsep{10pt}
    \renewcommand{\arraystretch}{1.1}
    \centering
    \caption{Properties of the CFHTWIR-Oph 98 AB system}
    \footnotesize
    \begin{tabular}{l c c c }
    \hline \hline
        Parameter  & Oph 98 A & Oph 98 B & Reference \\ \hline \hline
        \multicolumn{2}{l}{ASTROMETRY} \\ \hline 
        $\alpha$ (ICRS J2000.0) & \multicolumn{2}{c}{$16^\mathrm{h}27^\mathrm{m}44\fs226$} & \citet{2MASS} \\
        $\delta\,$ (ICRS J2000.0) & \multicolumn{2}{c}{$-23^\circ58'52\farcs14$} & \citet{2MASS} \\
        $\mu_{\alpha*}$ [mas yr$^{-1}$] & \multicolumn{2}{c}{$-7.2\pm2.0$} & \citet{Canovas2019} \\
        $\mu_{\delta}$ [mas yr$^{-1}$] & \multicolumn{2}{c}{$-25.5\pm1.7$} & \citet{Canovas2019} \\
        Parallax [mas] & \multicolumn{2}{c}{$7.29\pm0.22$} & \citet{Ortiz-Leon2018} \\
        Distance [pc] & \multicolumn{2}{c}{$137\pm4$} & \citet{Ortiz-Leon2018} \\
        \hline \hline
        \multicolumn{2}{l}{PHOTOMETRY} \\ \hline 
        2MASS \textit{J} [mag] & \multicolumn{2}{c}{$16.775\pm0.176$} & \citet{2MASS} \\
        2MASS \textit{H} [mag] & \multicolumn{2}{c}{$15.574\pm0.109$} & \citet{2MASS} \\
        2MASS \textit{Ks} [mag] & \multicolumn{2}{c}{$14.593\pm0.098$} & \citet{2MASS} \\
        F850LP [mag] & $19.696\pm0.018$ & $22.479\pm0.170$ & This paper \\ 
        F127M [mag] & $16.835\pm0.005$ & $19.027\pm0.013$ & This paper \\
        F139M [mag] & $16.959\pm0.005$ & $18.991\pm0.016$ & This paper \\
        WIRCAM \textit{J} [mag] & $17.015\pm0.016$ & $19.109\pm0.050$ & This paper \\
        WIRCAM \textit{H} [mag] & $15.851\pm0.016$ & $17.541\pm0.030$ & This paper \\
        WIRCAM \textit{Ks} [mag] & $14.917\pm0.009$ & $16.498\pm0.017$ & This paper \\ 
        WFCAM \textit{J} [mag] & $16.975\pm0.009$ & $19.042\pm0.038$ & This paper \\
        WFCAM \textit{H} [mag] & $15.826\pm0.013$ & $17.620\pm0.037$ & This paper \\
        WFCAM \textit{K} [mag] & $14.792\pm0.039$ & $16.408\pm0.047$ & This paper \\ 
        \hline \hline
        \multicolumn{2}{l}{FUNDAMENTAL PROPERTIES} \\ \hline
        $A_V$ [mag] & \multicolumn{2}{c}{$5\pm1$} & This paper \\
        Spectral type$^\dagger$ & M9--L1 & L2--L6 & This paper \\
        $\log(L_\mathrm{bol}/L_\odot)$ [dex] & $-2.85\pm0.06$ & $-3.49\pm0.06$ & This paper \\
        \hline
        & \multicolumn{2}{c}{1 Myr} \\
        \hline
        $T_\mathrm{eff}$ [K] & $2210\pm60$ & $1740\pm40$ & This paper \\
        log\,$g$ [dex] & $3.566^{+0.040}_{-0.048}$ & $3.436^{+0.010}_{-0.015}$ & This paper \\
        Radius [R$_\mathrm{Jup}$] & $2.61\pm0.05$ & $2.00^{+0.04}_{-0.03}$ & This paper \\
        Mass [M$_\mathrm{Jup}$] & $9.6\pm1.4$ & $4.1^{+0.4}_{-0.3}$ & This paper \\
        \hline
        & \multicolumn{2}{c}{3 Myr} \\
        \hline
        $T_\mathrm{eff}$ [K] & $2320\pm40$ & $1800\pm40$ & This paper \\
        log\,$g$ [dex] & $3.845^{+0.007}_{-0.008}$ & $3.748^{+0.015}_{-0.016}$ & This paper \\
        Radius [R$_\mathrm{Jup}$] & $2.38^{+0.07}_{-0.08}$ & $1.86\pm0.05$ & This paper \\
        Mass [M$_\mathrm{Jup}$] & $15.4\pm0.8$ & $7.8^{+0.7}_{-0.8}$ & This paper \\
        \hline
        & \multicolumn{2}{c}{7 Myr} \\
        \hline
        $T_\mathrm{eff}$ [K] & $2370\pm40$ & $1850^{+50}_{-40}$ & This paper \\
        log\,$g$ [dex] & $3.974^{+0.010}_{-0.008}$ & $3.984^{+0.002}_{-0.009}$ & This paper \\
        Radius [R$_\mathrm{Jup}$] & $2.27\pm0.07$ & $1.77^{+0.03}_{-0.05}$ & This paper \\
        Mass [M$_\mathrm{Jup}$] & $18.4^{+0.8}_{-0.7}$ & $11.6^{+0.4}_{-0.8}$ & This paper \\
        \hline \hline
        \multicolumn{2}{l}{BINARY CHARACTERISTICS} \\ \hline
        Separation [arcsec] & \multicolumn{2}{c}{$1.46\pm0.01$} & This paper \\
        Separation [au] & \multicolumn{2}{c}{$200\pm6$} & This paper \\
        Orbital period [yr]$^*$ & \multicolumn{2}{c}{$22\,000\pm1\,300$} & This paper \\
        Mass ratio$^*$ & \multicolumn{2}{c}{$0.509^{+0.017}_{-0.023}$} & This paper \\
        $E_b$ [$10^{39}$ erg]$^*$& \multicolumn{2}{c}{$8.8\pm1.4$} & This paper \\
        \hline
        \multicolumn{4}{l}{\footnotesize $^\dagger$Photometric estimates based on near-infrared \textsc{vl-g} spectral templates.}\\[-0.15cm]
        \multicolumn{4}{l}{\footnotesize $^*$Quantities calculated for a median age of 3 Myr.}
    \end{tabular}
    \label{t:properties}
\end{table*}

We calculated the luminosities of Oph~98~A and B using extinction-corrected MKO \textit{Ks} photometry, the \textit{K}-band bolometric corrections for young brown dwarfs from \citet{Filippazzo2015}, and a parallax for L1688 of $7.29\pm0.22$~mas \citep{Ortiz-Leon2018}.  We used a Monte Carlo approach to determine uncertainties, taking a uniform distribution of $A_V$ and spectral type, and normally-distributed uncertainties for the bolometric corrections and parallax.  We determine luminosities of $\log{(L_{\mathrm{bol}}/L_\odot)}=-2.85\pm0.06$~dex and $-3.49\pm0.06$~dex for Oph~98~A and B, respectively.

Table~\ref{t:properties} presents the masses, effective temperatures ($T_\mathrm{eff}$), radii, and surface gravities (log\,$g$) of Oph~98~A and B calculated from the DUSTY model isochrones of \citet{Chabrier2000} at ages of 1, 3 and 7~Myr. We estimate parameters for the bounds of the interquartile age range as the underlying age distribution of L1688 is unknown. Comparison of the luminosities to evolutionary models at the adopted age of 3~Myr yields masses of $15.4\pm0.8$ and $7.8^{+0.7}_{-0.8}$~$\mathrm{M_{Jup}}$ for Oph~98~A and B, respectively. Over the possible ages of the system (1--7~Myr), the mass of Oph~98~A could range from 9.6 to 18.4~$\mathrm{M_{Jup}}$, and the mass of Oph~98~B from 4.1 to 11.6~$\mathrm{M_{Jup}}$. The primary mass could therefore lie on either side of the planet/brown dwarf boundary ($\sim$13~$\mathrm{M_{Jup}}$), while the secondary is confidently in the planetary-mass regime for all plausible system ages. Calculated masses using the evolutionary models of \citet{Saumon2008} ($f_{\mathrm{sed}} = 2$) and \citet{Burrows1997} were found to agree with the DUSTY results to within the uncertainties.

From the model-derived masses at 3~Myr and measured angular separation, we also computed the binary mass ratio, orbital period and gravitational binding energy of the system, reported in Table~\ref{t:properties}. We used a median correction factor of 1.1 from projected separation to true semi-major axis \citep{DupuyLiu2011}.

\section{Discussion and Conclusions} \label{sec:discussion} 




\subsection{Comparison to Other Systems} 
\label{comparison}

With a primary mass near the deuterium-burning limit ($\sim$11--16~M$_\mathrm{Jup}$ depending on metallicity; \citealp{Spiegel2011}) and a companion inside the planetary mass range, Oph~98~AB is among the lowest-mass binaries known to date.
Only a handful of systems potentially made of two planetary-mass components have been discovered so far: the TW~Hya candidate member 2MASS~J1119$-$1137 \citep{Best2017}, the WISE~J1355$-$8258 spectral binary candidate in the AB~Dor moving group \citep{Theissen2020}, and the old field systems WISE~J0146+4234 \citep{Dupuy2015} and CFBDSIR~J1458+1013 \citep{Liu2011} that may be more massive depending on their ages.

Oph~98 is distinct from these binaries in three important aspects: the system's age, mass ratio and separation. The extremely young age of Ophiuchus makes Oph~98 the youngest system currently known with both components near or below the deuterium burning mass boundary, and is thus the only example of such a binary detected right after birth. Additionally, most of these planetary-mass binaries are in nearly equal-mass configurations. In contrast, Oph~98 has a significantly lower mass ratio ($q=0.509^{+0.017}_{-0.023}$) for a comparable total mass. Finally, all the systems listed above have very tight orbital separations ($<5$~au), while Oph~98~AB is on a considerably wider 200-au orbit. No old field binaries are known on such large separations in this mass regime (see \citealp{Fontanive2018}).

\begin{figure}
    \centering
    \includegraphics[width=0.45\textwidth]{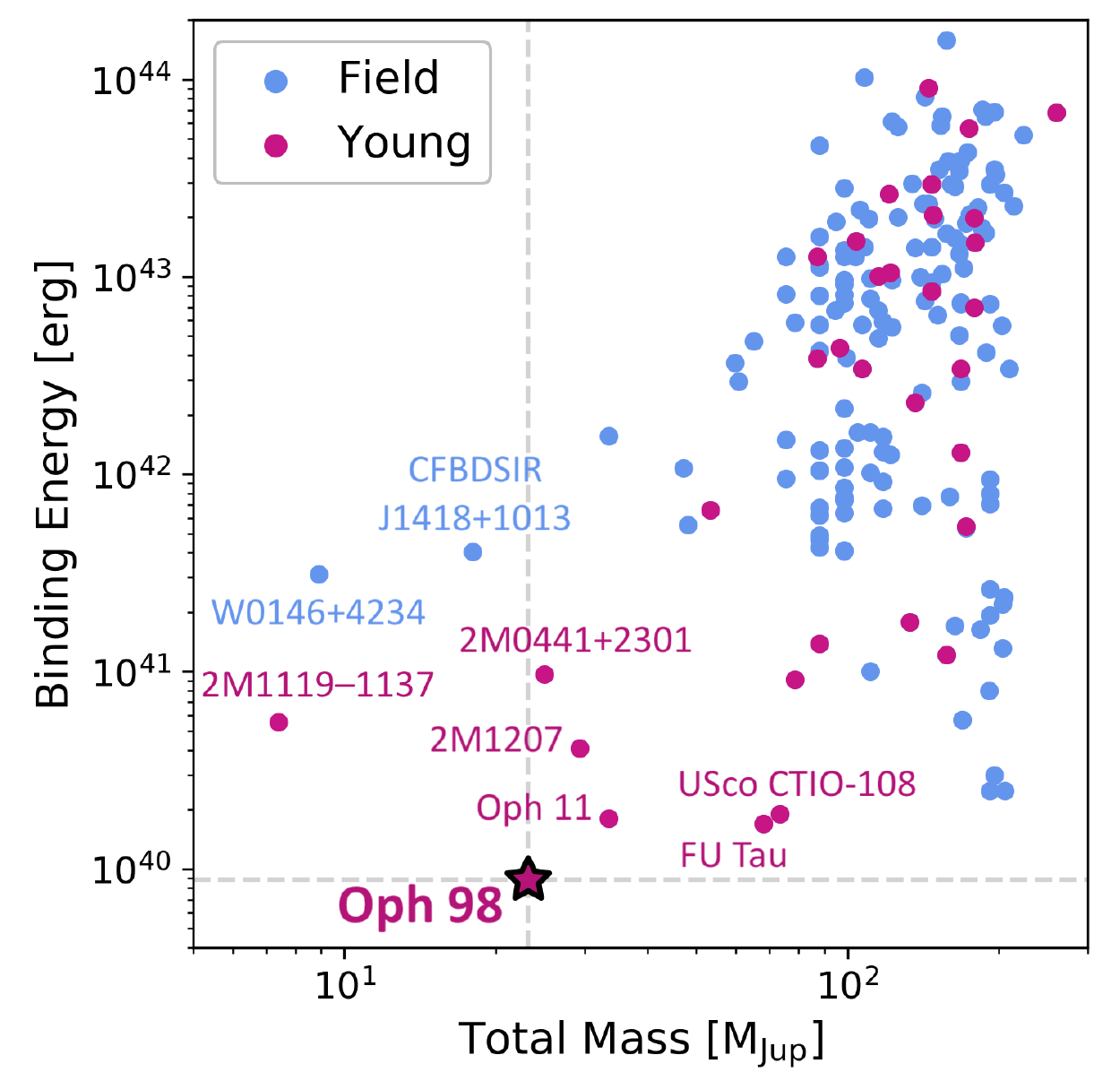}
    \caption{Binding energy plotted against total system mass for low-mass binaries in the field (blue) and young associations (magenta), based on the compilation from \citet{Faherty2020}. Oph~98 (magenta star) is among the lowest mass binaries currently known (vertical line) and has the weakest binding energy of any known system (horizontal line).}
    \label{f:Eb}
\end{figure}

In these aspects, Oph~98 is more akin to wide systems with very low-mass companions identified in young associations ($\leq10$~Myr). The closest analog to Oph~98 is certainly the other Ophiuchus brown dwarf binary, Oph~11 (17+14~M$_\mathrm{Jup}$, 240~au; \citealp{Close2007}). Lower mass ratio counterparts include 2M1207 \citep{Chauvin2005} and 2MASS~J0441+2301 \citep{Todorov2010}, with $\sim$20-M$_\mathrm{Jup}$ primaries and planetary-mass secondaries (4--5~M$_\mathrm{Jup}$), but significantly shorter separations of a few tens of au; or FU~Tau \citep{Luhman2009} and USco~CTIO-108 \citep{Bejar2008}, on separations of hundreds of au, but considerably higher-mass primaries ($\sim$50~M$_\mathrm{Jup}$) and secondaries around 15~M$_\mathrm{Jup}$. 

The combination for Oph~98~AB of small estimated component masses (15.4+7.8~M$_\mathrm{Jup}$) and large measured separation (200~au) results in a remarkably low binding energy of $E_\mathrm{b}=8.8\times10^{39}$~erg. This is lower by a factor of at least 2 than any of the binaries mentioned above, as illustrated in Figure~\ref{f:Eb}, with a second weakest binding energy belonging to FU~Tau (see compilation in \citealp{Faherty2020}). The Oph~98 system is therefore the brown dwarf binary with the lowest gravitational binding energy discovered to date.

\subsection{Formation Mechanisms}
\label{formation}

Oph~98, like each of the young low-mass binaries mentioned in \S~\ref{comparison}, provides a valuable example of a young system in an extreme configuration, offering key insight into formation for the very lowest-mass brown dwarfs. Indeed, as multiplicity is a direct outcome of formation, the properties of very young binaries can serve as key diagnostics of formation pathways. The youth of the Oph~98 system indicates that the involved mechanisms must operate on short Myr-level timescales, compatible with the rapid formation expected from the fragmentation of cloud cores or gravitational instability in disks. The weakly-bound nature of the system argues against violent dynamical processes that would have disrupted the binary, like the premature ejection of sub-stellar embryos from the natal cloud \citep{ReipurthClarke2001}. Likewise, formation and subsequent ejection of both components from the disk of a more massive star \citep{Stamatellos2007} seems implausible based on the binary configuration.

It is more probable that Oph~98~A formed in a star-like manner, through the fragmentation of molecular cloud cores \citep{Whitworth2007}. The low-mass companion Oph~98~B could have formed in the same way, or in the disk of the primary. The latter scenario is unlikely given the mass of the primary, too low to host a disk more massive than $\sim$1~M$_\mathrm{Jup}$, and the wide binary separation, compared to radii of $<\!30$--100~au for brown dwarf disks \citep{Testi2016}. The mass ratio of $\sim$0.5 also suggests a binary-like architecture rather than a planet-like origin for the secondary \citep{Lodato2005}, further supporting the hypothesis that Oph~98~AB emerged from a stellar formation process.

With a lower mass limit for brown dwarfs around $\sim$3~M$_\mathrm{Jup}$ in a star formation framework \citep{Whitworth2007} -- the minimum mass for opacity-limited fragmentation in turbulent cloud cores \citep{Silk1977} -- Oph~98~B thus extends the growing number of very young, planetary-mass objects populating the low-end tail of the star formation product (e.g. \citealp{Liu2013,Gagne2014,Gagne2015,Schneider2016}). The existence of such wide, very low-mass binaries arising from a stellar formation pathway is predicted in numerical simulations \citep{Bate2012}, although expected to be of rare occurrence.
As the binary with the weakest gravitational binding energy discovered to date, the Oph~98 system therefore represents an unmatched example of the extreme multiplicity outcome of stellar formation.

\acknowledgments

We thank the anonymous referee for a thorough and constructive report. C.F. acknowledges financial support from the Center for Space and Habitability (CSH). This work has been carried out within the framework of the NCCR PlanetS supported by the Swiss National Science Foundation.
Based on observations made with the NASA/ESA Hubble Space Telescope, which is operated by the Association of Universities for Research in Astronomy, Inc., under NASA contract NAS5-26555. These observations are associated with program \#12944. 
Support for this work was provided by NASA through grant numbers 12944, 14686, and 15201 from the Space Telescope Science Institute.
This research has been funded in part by grants from the Gordon and Betty Moore Foundation (grant GBMF8550) and the National Science Foundation (AST-1518339) awarded to M. Liu.
Based on observations obtained with WIRCam, a joint project of CFHT, Taiwan, Korea, Canada, France, at the Canada-France-Hawaii Telescope (CFHT) which is operated by the National Research Council (NRC) of Canada, the Institut National des Sciences de l'Univers of the Centre National de la Recherche Scientifique of France, and the University of Hawaii.
This research has benefited from the SpeX Prism Spectral Libraries, maintained by Adam Burgasser at \url{http://pono.ucsd.edu/~adam/browndwarfs/spexprism/}.
This work made use of the Database of Ultracool Parallaxes maintained by Trent Dupuy at \url{http://www.as.utexas.edu/~tdupuy/plx/Database_of_Ultracool_Parallaxes.html}.


\begin{thebibliography}{}
\expandafter\ifx\csname natexlab\endcsname\relax\def\natexlab#1{#1}\fi
\providecommand{\url}[1]{\href{#1}{#1}}
\providecommand{\dodoi}[1]{doi:~\href{http://doi.org/#1}{\nolinkurl{#1}}}
\providecommand{\doeprint}[1]{\href{http://ascl.net/#1}{\nolinkurl{http://ascl.net/#1}}}
\providecommand{\doarXiv}[1]{\href{https://arxiv.org/abs/#1}{\nolinkurl{https://arxiv.org/abs/#1}}}

\bibitem[{{Allers} \& {Liu}(2013)}]{AllersLiu2013}
{Allers}, K.~N., \& {Liu}, M.~C. 2013, \apj, 772, 79

\bibitem[{{Allers} \& {Liu}(2020)}]{Allers2020}
---. 2020, \pasp, 132, 104401

\bibitem[{{Alves de Oliveira} {et~al.}(2012){Alves de Oliveira}, {Moraux},
  {Bouvier}, \& {Bouy}}]{AlvesdeOliveira2012}
{Alves de Oliveira}, C., {Moraux}, E., {Bouvier}, J., \& {Bouy}, H. 2012, \aap,
  539, A151

\bibitem[{{Barman} {et~al.}(2011){Barman}, {Macintosh}, {Konopacky}, \&
  {Marois}}]{Barman2011}
{Barman}, T.~S., {Macintosh}, B., {Konopacky}, Q.~M., \& {Marois}, C. 2011,
  \apjl, 735, L39

\bibitem[{{Bate}(2012)}]{Bate2012}
{Bate}, M.~R. 2012, \mnras, 419, 3115

\bibitem[{{B{\'e}jar} {et~al.}(2008){B{\'e}jar}, {Zapatero Osorio},
  {P{\'e}rez-Garrido}, {{\'A}lvarez}, {Mart{\'\i}n}, {Rebolo},
  {Vill{\'o}-P{\'e}rez}, \& {D{\'\i}az-S{\'a}nchez}}]{Bejar2008}
{B{\'e}jar}, V.~J.~S., {Zapatero Osorio}, M.~R., {P{\'e}rez-Garrido}, A.,
  {et~al.} 2008, \apjl, 673, L185

\bibitem[{{Best} {et~al.}(2017){Best}, {Liu}, {Dupuy}, \& {Magnier}}]{Best2017}
{Best}, W. M.~J., {Liu}, M.~C., {Dupuy}, T.~J., \& {Magnier}, E.~A. 2017,
  \apjl, 843, L4

\bibitem[{Bradley {et~al.}(2019)Bradley, Sip{\H o}cz, Robitaille, Tollerud,
  Vin{\'{\i}}cius, Deil, Barbary, G{\"u}nther, Cara, Busko, Conseil,
  Droettboom, Bostroem, Bray, Bratholm, Wilson, Craig, Barentsen, Pascual,
  Donath, Greco, Perren, Lim, \& Kerzendorf}]{Bradley2019}
Bradley, L., Sip{\H o}cz, B., Robitaille, T., {et~al.} 2019, astropy/photutils:
  v0.6

\bibitem[{{Burgasser} {et~al.}(2016){Burgasser}, {Aganze}, {Escala}, {Lopez},
  {Choban}, {Jin}, {Iyer}, {Tallis}, {Suarez}, \& {Sahi}}]{SPLAT}
{Burgasser}, A.~J., {Aganze}, C., {Escala}, I., {et~al.} 2016, in American
  Astronomical Society Meeting Abstracts \#227, Vol. 227, 434.08

\bibitem[{{Burrows} {et~al.}(1997){Burrows}, {Marley}, {Hubbard}, {Lunine},
  {Guillot}, {Saumon}, {Freedman}, {Sudarsky}, \& {Sharp}}]{Burrows1997}
{Burrows}, A., {Marley}, M., {Hubbard}, W.~B., {et~al.} 1997, \apj, 491, 856

\bibitem[{{C{\'a}novas} {et~al.}(2019){C{\'a}novas}, {Cantero}, {Cieza},
  {Bombrun}, {Lammers}, {Mer{\'\i}n}, {Mora}, {Ribas}, \&
  {Ru{\'\i}z-Rodr{\'\i}guez}}]{Canovas2019}
{C{\'a}novas}, H., {Cantero}, C., {Cieza}, L., {et~al.} 2019, \aap, 626, A80

\bibitem[{{Cardelli} {et~al.}(1989){Cardelli}, {Clayton}, \&
  {Mathis}}]{Cardelli1989}
{Cardelli}, J.~A., {Clayton}, G.~C., \& {Mathis}, J.~S. 1989, \apj, 345, 245

\bibitem[{{Chabrier} {et~al.}(2000){Chabrier}, {Baraffe}, {Allard}, \&
  {Hauschildt}}]{Chabrier2000}
{Chabrier}, G., {Baraffe}, I., {Allard}, F., \& {Hauschildt}, P. 2000, \apj,
  542, 464

\bibitem[{{Chauvin} {et~al.}(2005){Chauvin}, {Lagrange}, {Dumas}, {Zuckerman},
  {Mouillet}, {Song}, {Beuzit}, \& {Lowrance}}]{Chauvin2005}
{Chauvin}, G., {Lagrange}, A.~M., {Dumas}, C., {et~al.} 2005, \aap, 438, L25

\bibitem[{{Close} {et~al.}(2007){Close}, {Zuckerman}, {Song}, {Barman},
  {Marois}, {Rice}, {Siegler}, {Macintosh}, {Becklin}, {Campbell}, {Lyke},
  {Conrad}, \& {Le Mignant}}]{Close2007}
{Close}, L.~M., {Zuckerman}, B., {Song}, I., {et~al.} 2007, \apj, 660, 1492

\bibitem[{{Cruz} {et~al.}(2009){Cruz}, {Kirkpatrick}, \&
  {Burgasser}}]{Cruz2009}
{Cruz}, K.~L., {Kirkpatrick}, J.~D., \& {Burgasser}, A.~J. 2009, \aj, 137, 3345

\bibitem[{{Cutri} {et~al.}(2003){Cutri}, {Skrutskie}, {van Dyk}, {Beichman},
  {Carpenter}, {Chester}, {Cambresy}, {Evans}, {Fowler}, {Gizis}, {Howard},
  {Huchra}, {Jarrett}, {Kopan}, {Kirkpatrick}, {Light}, {Marsh}, {McCallon},
  {Schneider}, {Stiening}, {Sykes}, {Weinberg}, {Wheaton}, {Wheelock}, \&
  {Zacarias}}]{2MASS}
{Cutri}, R.~M., {Skrutskie}, M.~F., {van Dyk}, S., {et~al.} 2003, VizieR Online
  Data Catalog, II/246

\bibitem[{{Dupuy} \& {Kraus}(2013)}]{DupuyKraus2013}
{Dupuy}, T.~J., \& {Kraus}, A.~L. 2013, Science, 341, 1492

\bibitem[{{Dupuy} \& {Liu}(2011)}]{DupuyLiu2011}
{Dupuy}, T.~J., \& {Liu}, M.~C. 2011, \apj, 733, 122

\bibitem[{{Dupuy} \& {Liu}(2012)}]{DupuyLiu2012}
---. 2012, \apjs, 201, 19

\bibitem[{{Dupuy} {et~al.}(2015){Dupuy}, {Liu}, \& {Leggett}}]{Dupuy2015}
{Dupuy}, T.~J., {Liu}, M.~C., \& {Leggett}, S.~K. 2015, \apj, 803, 102

\bibitem[{{Dupuy} {et~al.}(2018){Dupuy}, {Liu}, {Allers}, {Biller}, {Kratter},
  {Mann}, {Shkolnik}, {Kraus}, \& {Best}}]{Dupuy2018}
{Dupuy}, T.~J., {Liu}, M.~C., {Allers}, K.~N., {et~al.} 2018, \aj, 156, 57

\bibitem[{{Esplin} \& {Luhman}(2020)}]{EsplinLuhman2020}
{Esplin}, T.~L., \& {Luhman}, K.~L. 2020, \aj, 159, 282

\bibitem[{{Faherty} {et~al.}(2009){Faherty}, {Burgasser}, {Cruz}, {Shara},
  {Walter}, \& {Gelino}}]{Faherty2009}
{Faherty}, J.~K., {Burgasser}, A.~J., {Cruz}, K.~L., {et~al.} 2009, \aj, 137, 1

\bibitem[{{Faherty} {et~al.}(2020){Faherty}, {Goodman}, {Caselden}, {Colin},
  {Kuchner}, {Meisner}, {Gagn{\'e}}, {Schneider}, {Gonzales}, {Bardalez
  Gagliuffi}, {Logsdon}, {Allers}, \& {Burgasser}}]{Faherty2020}
{Faherty}, J.~K., {Goodman}, S., {Caselden}, D., {et~al.} 2020, \apj, 889, 176

\bibitem[{{Filippazzo} {et~al.}(2015){Filippazzo}, {Rice}, {Faherty}, {Cruz},
  {Van Gordon}, \& {Looper}}]{Filippazzo2015}
{Filippazzo}, J.~C., {Rice}, E.~L., {Faherty}, J., {et~al.} 2015, \apj, 810,
  158

\bibitem[{{Fontanive} {et~al.}(2018){Fontanive}, {Biller}, {Bonavita}, \&
  {Allers}}]{Fontanive2018}
{Fontanive}, C., {Biller}, B., {Bonavita}, M., \& {Allers}, K. 2018, \mnras,
  479, 2702

\bibitem[{{Fruchter} \& {Hook}(2002)}]{FruchterHook2002}
{Fruchter}, A.~S., \& {Hook}, R.~N. 2002, \pasp, 114, 144

\bibitem[{{Gagn{\'e}} {et~al.}(2014){Gagn{\'e}}, {Lafreni{\`e}re}, {Doyon},
  {Malo}, \& {Artigau}}]{Gagne2014}
{Gagn{\'e}}, J., {Lafreni{\`e}re}, D., {Doyon}, R., {Malo}, L., \& {Artigau},
  {\'E}. 2014, \apj, 783, 121

\bibitem[{{Gagn{\'e}} {et~al.}(2015){Gagn{\'e}}, {Faherty}, {Cruz},
  {Lafreni{\'e}re}, {Doyon}, {Malo}, {Burgasser}, {Naud}, {Artigau},
  {Bouchard}, {Gizis}, \& {Albert}}]{Gagne2015}
{Gagn{\'e}}, J., {Faherty}, J.~K., {Cruz}, K.~L., {et~al.} 2015, \apjs, 219, 33

\bibitem[{{Gaia Collaboration} {et~al.}(2016){Gaia Collaboration}, {Prusti},
  {de Bruijne}, {Brown}, {Vallenari}, {Babusiaux}, {Bailer-Jones}, {Bastian},
  {Biermann}, {Evans}, {Eyer}, {Jansen}, {Jordi}, {Klioner}, {Lammers},
  {Lindegren}, {Luri}, {Mignard}, {Milligan}, {Panem}, {Poinsignon},
  {Pourbaix}, {Randich}, {Sarri}, {Sartoretti}, {Siddiqui}, {Soubiran},
  {Valette}, {van Leeuwen}, {Walton}, {Aerts}, {Arenou}, {Cropper}, {Drimmel},
  {H{\o}g}, {Katz}, {Lattanzi}, {O'Mullane}, {Grebel}, {Holland}, {Huc},
  {Passot}, {Bramante}, {Cacciari}, {Casta{\~n}eda}, {Chaoul}, {Cheek}, {De
  Angeli}, {Fabricius}, {Guerra}, {Hern{\'a}ndez}, {Jean-Antoine-Piccolo},
  {Masana}, {Messineo}, {Mowlavi}, {Nienartowicz}, {Ord{\'o}{\~n}ez-Blanco},
  {Panuzzo}, {Portell}, {Richards}, {Riello}, {Seabroke}, {Tanga},
  {Th{\'e}venin}, {Torra}, {Els}, {Gracia-Abril}, {Comoretto},
  {Garcia-Reinaldos}, {Lock}, {Mercier}, {Altmann}, {Andrae}, {Astraatmadja},
  {Bellas-Velidis}, {Benson}, {Berthier}, {Blomme}, {Busso}, {Carry},
  {Cellino}, {Clementini}, {Cowell}, {Creevey}, {Cuypers}, {Davidson}, {De
  Ridder}, {de Torres}, {Delchambre}, {Dell'Oro}, {Ducourant}, {Fr{\'e}mat},
  {Garc{\'\i}a-Torres}, {Gosset}, {Halbwachs}, {Hambly}, {Harrison}, {Hauser},
  {Hestroffer}, {Hodgkin}, {Huckle}, {Hutton}, {Jasniewicz}, {Jordan},
  {Kontizas}, {Korn}, {Lanzafame}, {Manteiga}, {Moitinho}, {Muinonen},
  {Osinde}, {Pancino}, {Pauwels}, {Petit}, {Recio-Blanco}, {Robin}, {Sarro},
  {Siopis}, {Smith}, {Smith}, {Sozzetti}, {Thuillot}, {van Reeven}, {Viala},
  {Abbas}, {Abreu Aramburu}, {Accart}, {Aguado}, {Allan}, {Allasia},
  {Altavilla}, {{\'A}lvarez}, {Alves}, {Anderson}, {Andrei}, {Anglada Varela},
  {Antiche}, {Antoja}, {Ant{\'o}n}, {Arcay}, {Atzei}, {Ayache}, {Bach},
  {Baker}, {Balaguer-N{\'u}{\~n}ez}, {Barache}, {Barata}, {Barbier}, {Barblan},
  {Baroni}, {Barrado y Navascu{\'e}s}, {Barros}, {Barstow}, {Becciani},
  {Bellazzini}, {Bellei}, {Bello Garc{\'\i}a}, {Belokurov}, {Bendjoya},
  {Berihuete}, {Bianchi}, {Bienaym{\'e}}, {Billebaud}, {Blagorodnova},
  {Blanco-Cuaresma}, {Boch}, {Bombrun}, {Borrachero}, {Bouquillon}, {Bourda},
  {Bouy}, {Bragaglia}, {Breddels}, {Brouillet}, {Br{\"u}semeister},
  {Bucciarelli}, {Budnik}, {Burgess}, {Burgon}, {Burlacu}, {Busonero}, {Buzzi},
  {Caffau}, {Cambras}, {Campbell}, {Cancelliere}, {Cantat-Gaudin}, {Carlucci},
  {Carrasco}, {Castellani}, {Charlot}, {Charnas}, {Charvet}, {Chassat},
  {Chiavassa}, {Clotet}, {Cocozza}, {Collins}, {Collins}, {Costigan}, {Crifo},
  {Cross}, {Crosta}, {Crowley}, {Dafonte}, {Damerdji}, {Dapergolas}, {David},
  {David}, {De Cat}, {de Felice}, {de Laverny}, {De Luise}, {De March}, {de
  Martino}, {de Souza}, {Debosscher}, {del Pozo}, {Delbo}, {Delgado},
  {Delgado}, {di Marco}, {Di Matteo}, {Diakite}, {Distefano}, {Dolding}, {Dos
  Anjos}, {Drazinos}, {Dur{\'a}n}, {Dzigan}, {Ecale}, {Edvardsson}, {Enke},
  {Erdmann}, {Escolar}, {Espina}, {Evans}, {Eynard Bontemps}, {Fabre},
  {Fabrizio}, {Faigler}, {Falc{\~a}o}, {Farr{\`a}s Casas}, {Faye}, {Federici},
  {Fedorets}, {Fern{\'a}ndez-Hern{\'a}ndez}, {Fernique}, {Fienga}, {Figueras},
  {Filippi}, {Findeisen}, {Fonti}, {Fouesneau}, {Fraile}, {Fraser}, {Fuchs},
  {Furnell}, {Gai}, {Galleti}, {Galluccio}, {Garabato}, {Garc{\'\i}a-Sedano},
  {Gar{\'e}}, {Garofalo}, {Garralda}, {Gavras}, {Gerssen}, {Geyer}, {Gilmore},
  {Girona}, {Giuffrida}, {Gomes}, {Gonz{\'a}lez-Marcos},
  {Gonz{\'a}lez-N{\'u}{\~n}ez}, {Gonz{\'a}lez-Vidal}, {Granvik}, {Guerrier},
  {Guillout}, {Guiraud}, {G{\'u}rpide}, {Guti{\'e}rrez-S{\'a}nchez}, {Guy},
  {Haigron}, {Hatzidimitriou}, {Haywood}, {Heiter}, {Helmi}, {Hobbs},
  {Hofmann}, {Holl}, {Holland }, {Hunt}, {Hypki}, {Icardi}, {Irwin}, {Jevardat
  de Fombelle}, {Jofr{\'e}}, {Jonker}, {Jorissen}, {Julbe}, {Karampelas},
  {Kochoska}, {Kohley}, {Kolenberg}, {Kontizas}, {Koposov}, {Kordopatis},
  {Koubsky}, {Kowalczyk}, {Krone-Martins}, {Kudryashova}, {Kull}, {Bachchan},
  {Lacoste-Seris}, {Lanza}, {Lavigne}, {Le Poncin-Lafitte}, {Lebreton},
  {Lebzelter}, {Leccia}, {Leclerc}, {Lecoeur-Taibi}, {Lemaitre}, {Lenhardt},
  {Leroux}, {Liao}, {Licata}, {Lindstr{\o}m}, {Lister}, {Livanou}, {Lobel},
  {L{\"o}ffler}, {L{\'o}pez}, {Lopez-Lozano}, {Lorenz}, {Loureiro},
  {MacDonald}, {Magalh{\~a}es Fernandes}, {Managau}, {Mann}, {Mantelet},
  {Marchal}, {Marchant}, {Marconi}, {Marie}, {Marinoni}, {Marrese},
  {Marschalk{\'o}}, {Marshall}, {Mart{\'\i}n-Fleitas}, {Martino}, {Mary},
  {Matijevi{\v{c}}}, {Mazeh}, {McMillan}, {Messina}, {Mestre}, {Michalik},
  {Millar}, {Miranda}, {Molina}, {Molinaro}, {Molinaro}, {Moln{\'a}r},
  {Moniez}, {Montegriffo}, {Monteiro}, {Mor}, {Mora}, {Morbidelli}, {Morel},
  {Morgenthaler}, {Morley}, {Morris}, {Mulone}, {Muraveva}, {Musella},
  {Narbonne}, {Nelemans}, {Nicastro}, {Noval}, {Ord{\'e}novic},
  {Ordieres-Mer{\'e}}, {Osborne}, {Pagani}, {Pagano}, {Pailler}, {Palacin},
  {Palaversa}, {Parsons}, {Paulsen}, {Pecoraro}, {Pedrosa}, {Pentik{\"a}inen},
  {Pereira}, {Pichon}, {Piersimoni}, {Pineau}, {Plachy}, {Plum}, {Poujoulet},
  {Pr{\v{s}}a}, {Pulone}, {Ragaini}, {Rago}, {Rambaux}, {Ramos-Lerate},
  {Ranalli}, {Rauw}, {Read}, {Regibo}, {Renk}, {Reyl{\'e}}, {Ribeiro},
  {Rimoldini}, {Ripepi}, {Riva}, {Rixon}, {Roelens}, {Romero-G{\'o}mez},
  {Rowell}, {Royer}, {Rudolph}, {Ruiz-Dern}, {Sadowski}, {Sagrist{\`a}
  Sell{\'e}s}, {Sahlmann}, {Salgado}, {Salguero}, {Sarasso}, {Savietto},
  {Schnorhk}, {Schultheis}, {Sciacca}, {Segol}, {Segovia}, {Segransan},
  {Serpell}, {Shih}, {Smareglia}, {Smart}, {Smith}, {Solano}, {Solitro},
  {Sordo}, {Soria Nieto}, {Souchay}, {Spagna}, {Spoto}, {Stampa}, {Steele},
  {Steidelm{\"u}ller}, {Stephenson}, {Stoev}, {Suess}, {S{\"u}veges}, {Surdej},
  {Szabados}, {Szegedi-Elek}, {Tapiador}, {Taris}, {Tauran}, {Taylor},
  {Teixeira}, {Terrett}, {Tingley}, {Trager}, {Turon}, {Ulla}, {Utrilla},
  {Valentini}, {van Elteren}, {Van Hemelryck}, {van Leeuwen}, {Varadi},
  {Vecchiato}, {Veljanoski}, {Via}, {Vicente}, {Vogt}, {Voss}, {Votruba},
  {Voutsinas}, {Walmsley}, {Weiler}, {Weingrill}, {Werner}, {Wevers},
  {Whitehead}, {Wyrzykowski}, {Yoldas}, {{\v{Z}}erjal}, {Zucker}, {Zurbach},
  {Zwitter}, {Alecu}, {Allen}, {Allende Prieto}, {Amorim},
  {Anglada-Escud{\'e}}, {Arsenijevic}, {Azaz}, {Balm}, {Beck}, {Bernstein},
  {Bigot}, {Bijaoui}, {Blasco}, {Bonfigli}, {Bono}, {Boudreault}, {Bressan},
  {Brown}, {Brunet}, {Bunclark}, {Buonanno}, {Butkevich}, {Carret}, {Carrion},
  {Chemin}, {Ch{\'e}reau}, {Corcione}, {Darmigny}, {de Boer}, {de Teodoro}, {de
  Zeeuw}, {Delle Luche}, {Domingues}, {Dubath}, {Fodor}, {Fr{\'e}zouls},
  {Fries}, {Fustes}, {Fyfe}, {Gallardo}, {Gallegos}, {Gardiol}, {Gebran},
  {Gomboc}, {G{\'o}mez}, {Grux}, {Gueguen}, {Heyrovsky}, {Hoar}, {Iannicola},
  {Isasi Parache}, {Janotto}, {Joliet}, {Jonckheere}, {Keil}, {Kim},
  {Klagyivik}, {Klar}, {Knude}, {Kochukhov}, {Kolka}, {Kos}, {Kutka}, {Lainey},
  {LeBouquin}, {Liu}, {Loreggia}, {Makarov}, {Marseille}, {Martayan},
  {Martinez-Rubi}, {Massart}, {Meynadier}, {Mignot}, {Munari}, {Nguyen},
  {Nordlander}, {Ocvirk}, {O'Flaherty}, {Olias Sanz}, {Ortiz}, {Osorio},
  {Oszkiewicz}, {Ouzounis}, {Palmer}, {Park}, {Pasquato}, {Peltzer}, {Peralta},
  {P{\'e}turaud}, {Pieniluoma}, {Pigozzi}, {Poels}, {Prat}, {Prod'homme},
  {Raison}, {Rebordao}, {Risquez}, {Rocca-Volmerange}, {Rosen}, {Ruiz-Fuertes},
  {Russo}, {Sembay}, {Serraller Vizcaino}, {Short}, {Siebert}, {Silva},
  {Sinachopoulos}, {Slezak}, {Soffel}, {Sosnowska}, {Strai{\v{z}}ys}, {ter
  Linden}, {Terrell}, {Theil}, {Tiede}, {Troisi}, {Tsalmantza}, {Tur},
  {Vaccari}, {Vachier}, {Valles}, {Van Hamme}, {Veltz}, {Virtanen}, {Wallut},
  {Wichmann}, {Wilkinson}, {Ziaeepour}, \& {Zschocke}}]{Gaia2016}
{Gaia Collaboration}, {Prusti}, T., {de Bruijne}, J.~H.~J., {et~al.} 2016,
  \aap, 595, A1

\bibitem[{{Gaia Collaboration} {et~al.}(2018){Gaia Collaboration}, {Brown},
  {Vallenari}, {Prusti}, {de Bruijne}, {Babusiaux}, {Bailer-Jones}, {Biermann},
  {Evans}, {Eyer}, {Jansen}, {Jordi}, {Klioner}, {Lammers}, {Lindegren},
  {Luri}, {Mignard}, {Panem}, {Pourbaix}, {Randich}, {Sartoretti}, {Siddiqui},
  {Soubiran}, {van Leeuwen}, {Walton}, {Arenou}, {Bastian}, {Cropper},
  {Drimmel}, {Katz}, {Lattanzi}, {Bakker}, {Cacciari}, {Casta{\~n}eda},
  {Chaoul}, {Cheek}, {De Angeli}, {Fabricius}, {Guerra}, {Holl}, {Masana},
  {Messineo}, {Mowlavi}, {Nienartowicz}, {Panuzzo}, {Portell}, {Riello},
  {Seabroke}, {Tanga}, {Th{\'e}venin}, {Gracia-Abril}, {Comoretto},
  {Garcia-Reinaldos}, {Teyssier}, {Altmann}, {Andrae}, {Audard},
  {Bellas-Velidis}, {Benson}, {Berthier}, {Blomme}, {Burgess}, {Busso},
  {Carry}, {Cellino}, {Clementini}, {Clotet}, {Creevey}, {Davidson}, {De
  Ridder}, {Delchambre}, {Dell'Oro}, {Ducourant},
  {Fern{\'a}ndez-Hern{\'a}ndez}, {Fouesneau}, {Fr{\'e}mat}, {Galluccio},
  {Garc{\'\i}a-Torres}, {Gonz{\'a}lez-N{\'u}{\~n}ez}, {Gonz{\'a}lez-Vidal},
  {Gosset}, {Guy}, {Halbwachs}, {Hambly}, {Harrison}, {Hern{\'a}ndez},
  {Hestroffer}, {Hodgkin}, {Hutton}, {Jasniewicz}, {Jean-Antoine-Piccolo},
  {Jordan}, {Korn}, {Krone-Martins}, {Lanzafame}, {Lebzelter}, {L{\"o}ffler},
  {Manteiga}, {Marrese}, {Mart{\'\i}n-Fleitas}, {Moitinho}, {Mora}, {Muinonen},
  {Osinde}, {Pancino}, {Pauwels}, {Petit}, {Recio-Blanco}, {Richards},
  {Rimoldini}, {Robin}, {Sarro}, {Siopis}, {Smith}, {Sozzetti}, {S{\"u}veges},
  {Torra}, {van Reeven}, {Abbas}, {Abreu Aramburu}, {Accart}, {Aerts},
  {Altavilla}, {{\'A}lvarez}, {Alvarez}, {Alves}, {Anderson}, {Andrei},
  {Anglada Varela}, {Antiche}, {Antoja}, {Arcay}, {Astraatmadja}, {Bach},
  {Baker}, {Balaguer-N{\'u}{\~n}ez}, {Balm}, {Barache}, {Barata}, {Barbato},
  {Barblan}, {Barklem}, {Barrado}, {Barros}, {Barstow}, {Bartholom{\'e}
  Mu{\~n}oz}, {Bassilana}, {Becciani}, {Bellazzini}, {Berihuete}, {Bertone},
  {Bianchi}, {Bienaym{\'e}}, {Blanco-Cuaresma}, {Boch}, {Boeche}, {Bombrun},
  {Borrachero}, {Bossini}, {Bouquillon}, {Bourda}, {Bragaglia}, {Bramante},
  {Breddels}, {Bressan}, {Brouillet}, {Br{\"u}semeister}, {Brugaletta},
  {Bucciarelli}, {Burlacu}, {Busonero}, {Butkevich}, {Buzzi}, {Caffau},
  {Cancelliere}, {Cannizzaro}, {Cantat-Gaudin}, {Carballo}, {Carlucci},
  {Carrasco}, {Casamiquela}, {Castellani}, {Castro-Ginard}, {Charlot},
  {Chemin}, {Chiavassa}, {Cocozza}, {Costigan}, {Cowell}, {Crifo}, {Crosta},
  {Crowley}, {Cuypers}, {Dafonte}, {Damerdji}, {Dapergolas}, {David}, {David},
  {de Laverny}, {De Luise}, {De March}, {de Martino}, {de Souza}, {de Torres},
  {Debosscher}, {del Pozo}, {Delbo}, {Delgado}, {Delgado}, {Di Matteo},
  {Diakite}, {Diener}, {Distefano}, {Dolding}, {Drazinos}, {Dur{\'a}n},
  {Edvardsson}, {Enke}, {Eriksson}, {Esquej}, {Eynard Bontemps}, {Fabre},
  {Fabrizio}, {Faigler}, {Falc{\~a}o}, {Farr{\`a}s Casas}, {Federici},
  {Fedorets}, {Fernique}, {Figueras}, {Filippi}, {Findeisen}, {Fonti},
  {Fraile}, {Fraser}, {Fr{\'e}zouls}, {Gai}, {Galleti}, {Garabato},
  {Garc{\'\i}a-Sedano}, {Garofalo}, {Garralda}, {Gavel}, {Gavras}, {Gerssen},
  {Geyer}, {Giacobbe}, {Gilmore}, {Girona}, {Giuffrida}, {Glass}, {Gomes},
  {Granvik}, {Gueguen}, {Guerrier}, {Guiraud}, {Guti{\'e}rrez-S{\'a}nchez},
  {Haigron}, {Hatzidimitriou}, {Hauser}, {Haywood}, {Heiter}, {Helmi}, {Heu},
  {Hilger}, {Hobbs}, {Hofmann}, {Holland}, {Huckle}, {Hypki}, {Icardi},
  {Jan{\ss}en}, {Jevardat de Fombelle}, {Jonker}, {Juh{\'a}sz}, {Julbe},
  {Karampelas}, {Kewley}, {Klar}, {Kochoska}, {Kohley}, {Kolenberg},
  {Kontizas}, {Kontizas}, {Koposov}, {Kordopatis}, {Kostrzewa-Rutkowska},
  {Koubsky}, {Lambert}, {Lanza}, {Lasne}, {Lavigne}, {Le Fustec}, {Le
  Poncin-Lafitte}, {Lebreton}, {Leccia}, {Leclerc}, {Lecoeur-Taibi},
  {Lenhardt}, {Leroux}, {Liao}, {Licata}, {Lindstr{\o}m}, {Lister}, {Livanou},
  {Lobel}, {L{\'o}pez}, {Managau}, {Mann}, {Mantelet}, {Marchal}, {Marchant},
  {Marconi}, {Marinoni}, {Marschalk{\'o}}, {Marshall}, {Martino}, {Marton},
  {Mary}, {Massari}, {Matijevi{\v{c}}}, {Mazeh}, {McMillan}, {Messina},
  {Michalik}, {Millar}, {Molina}, {Molinaro}, {Moln{\'a}r}, {Montegriffo},
  {Mor}, {Morbidelli}, {Morel}, {Morris}, {Mulone}, {Muraveva}, {Musella},
  {Nelemans}, {Nicastro}, {Noval}, {O'Mullane}, {Ord{\'e}novic},
  {Ord{\'o}{\~n}ez-Blanco}, {Osborne}, {Pagani}, {Pagano}, {Pailler},
  {Palacin}, {Palaversa}, {Panahi}, {Pawlak}, {Piersimoni}, {Pineau}, {Plachy},
  {Plum}, {Poggio}, {Poujoulet}, {Pr{\v{s}}a}, {Pulone}, {Racero}, {Ragaini},
  {Rambaux}, {Ramos-Lerate}, {Regibo}, {Reyl{\'e}}, {Riclet}, {Ripepi}, {Riva},
  {Rivard}, {Rixon}, {Roegiers}, {Roelens}, {Romero-G{\'o}mez}, {Rowell},
  {Royer}, {Ruiz-Dern}, {Sadowski}, {Sagrist{\`a} Sell{\'e}s}, {Sahlmann},
  {Salgado}, {Salguero}, {Sanna}, {Santana-Ros}, {Sarasso}, {Savietto},
  {Schultheis}, {Sciacca}, {Segol}, {Segovia}, {S{\'e}gransan}, {Shih},
  {Siltala}, {Silva}, {Smart}, {Smith}, {Solano}, {Solitro}, {Sordo}, {Soria
  Nieto}, {Souchay}, {Spagna}, {Spoto}, {Stampa}, {Steele},
  {Steidelm{\"u}ller}, {Stephenson}, {Stoev}, {Suess}, {Surdej}, {Szabados},
  {Szegedi-Elek}, {Tapiador}, {Taris}, {Tauran}, {Taylor}, {Teixeira},
  {Terrett}, {Teyssand ier}, {Thuillot}, {Titarenko}, {Torra Clotet}, {Turon},
  {Ulla}, {Utrilla}, {Uzzi}, {Vaillant}, {Valentini}, {Valette}, {van Elteren},
  {Van Hemelryck}, {van Leeuwen}, {Vaschetto}, {Vecchiato}, {Veljanoski},
  {Viala}, {Vicente}, {Vogt}, {von Essen}, {Voss}, {Votruba}, {Voutsinas},
  {Walmsley}, {Weiler}, {Wertz}, {Wevers}, {Wyrzykowski}, {Yoldas},
  {{\v{Z}}erjal}, {Ziaeepour}, {Zorec}, {Zschocke}, {Zucker}, {Zurbach}, \&
  {Zwitter}}]{Gaia2018}
{Gaia Collaboration}, {Brown}, A.~G.~A., {Vallenari}, A., {et~al.} 2018, \aap,
  616, A1

\bibitem[{{King} {et~al.}(2012){King}, {Parker}, {Patience}, \&
  {Goodwin}}]{King2012}
{King}, R.~R., {Parker}, R.~J., {Patience}, J., \& {Goodwin}, S.~P. 2012,
  \mnras, 421, 2025

\bibitem[{{Kroupa} {et~al.}(2013){Kroupa}, {Weidner}, {Pflamm-Altenburg},
  {Thies}, {Dabringhausen}, {Marks}, \& {Maschberger}}]{Kroupa2013}
{Kroupa}, P., {Weidner}, C., {Pflamm-Altenburg}, J., {et~al.} 2013, {The
  Stellar and Sub-Stellar Initial Mass Function of Simple and Composite
  Populations}, Vol.~5 (Springer Science), 115

\bibitem[{{Landsman}(1993)}]{IDLAstro}
{Landsman}, W.~B. 1993, in Astronomical Society of the Pacific Conference
  Series, Vol.~52, Astronomical Data Analysis Software and Systems II, ed.
  R.~J. {Hanisch}, R.~J.~V. {Brissenden}, \& J.~{Barnes}, 246

\bibitem[{{Liu} {et~al.}(2016){Liu}, {Dupuy}, \& {Allers}}]{Liu2016}
{Liu}, M.~C., {Dupuy}, T.~J., \& {Allers}, K.~N. 2016, \apj, 833, 96

\bibitem[{{Liu} {et~al.}(2011){Liu}, {Delorme}, {Dupuy}, {Bowler}, {Albert},
  {Artigau}, {Reyl{\'e}}, {Forveille}, \& {Delfosse}}]{Liu2011}
{Liu}, M.~C., {Delorme}, P., {Dupuy}, T.~J., {et~al.} 2011, \apj, 740, 108

\bibitem[{{Liu} {et~al.}(2013){Liu}, {Magnier}, {Deacon}, {Allers}, {Dupuy},
  {Kotson}, {Aller}, {Burgett}, {Chambers}, {Draper}, {Hodapp}, {Jedicke},
  {Kaiser}, {Kudritzki}, {Metcalfe}, {Morgan}, {Price}, {Tonry}, \&
  {Wainscoat}}]{Liu2013}
{Liu}, M.~C., {Magnier}, E.~A., {Deacon}, N.~R., {et~al.} 2013, \apjl, 777, L20

\bibitem[{{Lodato} {et~al.}(2005){Lodato}, {Delgado-Donate}, \&
  {Clarke}}]{Lodato2005}
{Lodato}, G., {Delgado-Donate}, E., \& {Clarke}, C.~J. 2005, \mnras, 364, L91

\bibitem[{{Luhman} {et~al.}(2009){Luhman}, {Mamajek}, {Allen}, {Muench}, \&
  {Finkbeiner}}]{Luhman2009}
{Luhman}, K.~L., {Mamajek}, E.~E., {Allen}, P.~R., {Muench}, A.~A., \&
  {Finkbeiner}, D.~P. 2009, \apj, 691, 1265

\bibitem[{{Luhman} {et~al.}(2017){Luhman}, {Mamajek}, {Shukla}, \&
  {Loutrel}}]{Luhman2017}
{Luhman}, K.~L., {Mamajek}, E.~E., {Shukla}, S.~J., \& {Loutrel}, N.~P. 2017,
  \aj, 153, 46

\bibitem[{{Marois} {et~al.}(2008){Marois}, {Macintosh}, {Barman}, {Zuckerman},
  {Song}, {Patience}, {Lafreni{\`e}re}, \& {Doyon}}]{Marois2008}
{Marois}, C., {Macintosh}, B., {Barman}, T., {et~al.} 2008, Science, 322, 1348

\bibitem[{{Mohanty} {et~al.}(2007){Mohanty}, {Jayawardhana}, {Hu{\'e}lamo}, \&
  {Mamajek}}]{Mohanty2007}
{Mohanty}, S., {Jayawardhana}, R., {Hu{\'e}lamo}, N., \& {Mamajek}, E. 2007,
  \apj, 657, 1064

\bibitem[{{Ortiz-Le{\'o}n} {et~al.}(2018){Ortiz-Le{\'o}n}, {Loinard}, {Dzib},
  {Kounkel}, {Galli}, {Tobin}, {Evans}, {Hartmann}, {Rodr{\'\i}guez},
  {Brice{\~n}o}, {Torres}, \& {Mioduszewski}}]{Ortiz-Leon2018}
{Ortiz-Le{\'o}n}, G.~N., {Loinard}, L., {Dzib}, S.~A., {et~al.} 2018, \apjl,
  869, L33

\bibitem[{{Reipurth} \& {Clarke}(2001)}]{ReipurthClarke2001}
{Reipurth}, B., \& {Clarke}, C. 2001, \aj, 122, 432

\bibitem[{{Saumon} \& {Marley}(2008)}]{Saumon2008}
{Saumon}, D., \& {Marley}, M.~S. 2008, \apj, 689, 1327

\bibitem[{{Schneider} {et~al.}(2016){Schneider}, {Windsor}, {Cushing},
  {Kirkpatrick}, \& {Wright}}]{Schneider2016}
{Schneider}, A.~C., {Windsor}, J., {Cushing}, M.~C., {Kirkpatrick}, J.~D., \&
  {Wright}, E.~L. 2016, \apjl, 822, L1

\bibitem[{{Silk}(1977)}]{Silk1977}
{Silk}, J. 1977, \apj, 214, 152

\bibitem[{{Spiegel} {et~al.}(2011){Spiegel}, {Burrows}, \&
  {Milsom}}]{Spiegel2011}
{Spiegel}, D.~S., {Burrows}, A., \& {Milsom}, J.~A. 2011, \apj, 727, 57

\bibitem[{{Stamatellos} {et~al.}(2007){Stamatellos}, {Hubber}, \&
  {Whitworth}}]{Stamatellos2007}
{Stamatellos}, D., {Hubber}, D.~A., \& {Whitworth}, A.~P. 2007, \mnras, 382,
  L30

\bibitem[{{Testi} {et~al.}(2016){Testi}, {Natta}, {Scholz}, {Tazzari}, {Ricci},
  \& {de Gregorio Monsalvo}}]{Testi2016}
{Testi}, L., {Natta}, A., {Scholz}, A., {et~al.} 2016, \aap, 593, A111

\bibitem[{{Theissen} {et~al.}(2020){Theissen}, {Bardalez Gagliuffi}, {Faherty},
  {Gagn{\'e}}, \& {Burgasser}}]{Theissen2020}
{Theissen}, C.~A., {Bardalez Gagliuffi}, D.~C., {Faherty}, J.~K., {Gagn{\'e}},
  J., \& {Burgasser}, A. 2020, Research Notes of the American Astronomical
  Society, 4, 67

\bibitem[{{Todorov} {et~al.}(2010){Todorov}, {Luhman}, \&
  {McLeod}}]{Todorov2010}
{Todorov}, K., {Luhman}, K.~L., \& {McLeod}, K.~K. 2010, \apjl, 714, L84

\bibitem[{{Whitworth} {et~al.}(2007){Whitworth}, {Bate}, {Nordlund},
  {Reipurth}, \& {Zinnecker}}]{Whitworth2007}
{Whitworth}, A., {Bate}, M.~R., {Nordlund}, {\r{A}}., {Reipurth}, B., \&
  {Zinnecker}, H. 2007, in Protostars and Planets V, ed. B.~{Reipurth},
  D.~{Jewitt}, \& K.~{Keil}, 459

\bibitem[{{Winters} {et~al.}(2015){Winters}, {Henry}, {Lurie}, {Hambly}, {Jao},
  {Bartlett}, {Boyd}, {Dieterich}, {Finch}, {Hosey}, {Ianna}, {Riedel},
  {Slatten}, \& {Subasavage}}]{Winters2015}
{Winters}, J.~G., {Henry}, T.~J., {Lurie}, J.~C., {et~al.} 2015, \aj, 149, 5

\end{thebibliography}
\bibliographystyle{aasjournal}

\end{document}